\documentclass[
 reprint,
superscriptaddress,
 amsmath,amssymb,
prl,
floatfix,
]{revtex4-2}

\usepackage{graphicx}
\usepackage{dcolumn}
\usepackage{bm}
\usepackage{color}
\usepackage[colorlinks,urlcolor=blue,citecolor=blue,linkcolor=blue]{hyperref}
\usepackage{url}

\draft 

\begin{document}


\title{Fabrication of high-quality PMMA/SiO$_x$ spaced planar microcavities for strong coupling of light with monolayer WS$_2$ excitons} 

\author{Tinghe~Yun}%
\affiliation{ARC Centre of Excellence in Future Low-Energy Electronics Technologies and Department of Quantum Science and Technology, Research School of Physics, The Australian National University, Canberra, ACT 2601, Australia}
\affiliation{Songshan Lake Materials Laboratory, Dongguan 523808, Guangdong, China} 
\affiliation{Institute of Physics, Chinese Academy of Science, Beijing, 100190, China}

\author{Eliezer~Estrecho}%
\affiliation{ARC Centre of Excellence in Future Low-Energy Electronics Technologies and Department of Quantum Science and Technology, Research School of Physics, The Australian National University, Canberra, ACT 2601, Australia}

\author{Andrew~G.~Truscott}%
\affiliation{Department of Quantum Science and Technology, Research School of Physics, The Australian National University, Canberra, ACT 2601, Australia}

\author{Elena~A.~Ostrovskaya}
\email{elena.ostrovskaya@anu.edu.au}
\affiliation{ARC Centre of Excellence in Future Low-Energy Electronics Technologies and Department of Quantum Science and Technology, Research School of Physics, The Australian National University, Canberra, ACT 2601, Australia}

\author{Matthias~J.~Wurdack}
\email{matthias.wurdack@anu.edu.au}
\affiliation{ARC Centre of Excellence in Future Low-Energy Electronics Technologies and Department of Quantum Science and Technology, Research School of Physics, The Australian National University, Canberra, ACT 2601, Australia}

\date{\today}

\begin{abstract}
Exciton polaritons in atomically-thin transition metal dichalcogenide crystals (monolayer TMDCs) have emerged as a promising candidate to enable topological transport, ultra-efficient laser technologies, and collective quantum phenomena such as polariton condensation and superfluidity at room temperature. However, integrating monolayer TMDCs into high-quality planar microcavities to achieve the required strong coupling between the cavity photons and the TMDC excitons (bound electron-hole pairs) has proven challenging.  Previous approaches to integration had to compromise between various adverse effects on the strength of light-matter interactions in the monolayer, the cavity photon lifetime,  and the lateral size of the microcavity.  Here,  we demonstrate a scalable approach to fabricating high-quality planar microcavities with an integrated monolayer WS$_2$ layer-by-layer by using  polymethyl methacrylate/silicon oxide (PMMA/SiO$_x$) as a cavity spacer. Because the exciton oscillator strength is well protected against the required processing steps by the PMMA layer, the microcavities investigated in this work, which have quality factors of above $10^3$, can operate in the strong light-matter coupling regime at room temperature. This is an important step towards fabricating wafer-scale and patterned microcavities for engineering the exciton-polariton potential landscape, which is essential for enabling many proposed technologies.
\end{abstract}


\maketitle 

Planar microcavities with embedded exciton-hosting materials based on the design of vertical-cavity surface-emitting lasers (VCSELs) have enabled the regime of strong light-matter coupling between excitons and photons \cite{Weissbuch1992}. The formation of exciton polaritons (polaritons herein), part-light part-matter bosonic quasiparticles, in these semiconductor microcavities has triggered significant research efforts, which led to the creation of Bose-Einstein condensates (BEC) \cite{Kasprzak2006} and superfluids \cite{Amo2009}, the demonstration of non-Hermitian effects \cite{Gao2015}, the development of electrically-driven ultra-efficient lasers \cite{Schneider2013} and the realisation of topological edge transport \cite{Klembt2018} on microchips. By using semiconductors with large exciton binding energies, some of these observations were successfully realised at room temperature \cite{Christopoulos2007,Guillet2011,Su2017,Dietrich2016,Dusel2021,Lerario2017,Su2021,Su2020}.

A promising family of inorganic semiconductors for future polaritonic devices are atomically-thin TMDCs \cite{Mak2010,Liu2014,Sanvitto2016,Schneider2018,Gu2019}. However, integrating these monolayers into planar microcavities to create polaritons has turned out to be challenging because they are extremely fragile and easily damaged by various fabrication techniques \cite{yun2021influence}. Many attempts at TMDC integration by using metallic top mirrors \cite{lundt2016room,Gu2019} or depositing the top structures directly on top of the monolayer with aggressive material deposition techniques \cite{Liu2014,Chen2017}, have led to dramatic degradation of the photon lifetimes, exciton quantum yields, or exciton-photon coupling strengths. These properties are significantly improved by recent techniques for integrating TMDCs into all-dielectric microcavities. However, these approaches are based on micrometer-sized distributed Bragg reflector (DBR) chips \cite{Lundt2019,Rupprecht2021,Wurdack2021,Solanas2021,Zhao2022} or mechanically exfoliated hexagonal Boron Nitride (hBN) crystals \cite{Knopf2019,Shan2021}, which massively limit its technological scalability. While all of these approaches have enabled the demonstration of many striking effects of TMDC polaritons \cite{Liu2014,Schneider2018}, which include signatures of bosonic condensation \cite{Waldherr2018,Solanas2021}, the spin-valley Hall effect \cite{Lundt2019}, and ballistic propagation at room temperature \cite{Wurdack2021}, fabrication recipes which allow for both scalability and high performance of the final device to enable future TMDC polariton based technologies are still to be developed.

In this work, we introduce a method for integrating monolayer WS$_2$ into high-quality planar microcavities with the lateral size mainly limited by the size of the monolayer. In particular, we layer PMMA/SiO$_x$ on top of the monolayer before depositing the top DBR via plasma enhanced chemical vapour deposition (PECVD). This is in contrast to using metal as the top mirror of the cavity as was done for the heavily investigated  DBR/metal cavities \cite{lundt2016room,Lundt2017,Wurdack2017,Gu2019,Gu2021,Datta2021}. We study the effects of each processing step on the optical properties of the WS$_2$ monolayer showing that the exciton oscillator strength is well preserved inside these structures, which enables strong light-matter coupling at room temperature. Finally, we demonstrate that the quality factors (Q-factors) of these microcavities can reach values on the order of $10^3$, therefore  exceeding the Q-factors of DBR/metal cavities by $1-2$ orders of magnitude \cite{lundt2016room,Lundt2017,Wurdack2017,Datta2021} (see Supplementary Table S1). TMDC based microcavities with comparable Q-factors were previously only realised  on microchips by using the technologies that inherently limit the lateral size of the devices \cite{Knopf2019,Lundt2019,Rupprecht2021,Wurdack2021} or dramatically strain the monolayer \cite{Zhao2021,Shan2021}.


\begin{figure}[h]
\centering
\includegraphics[width=8.6 cm]{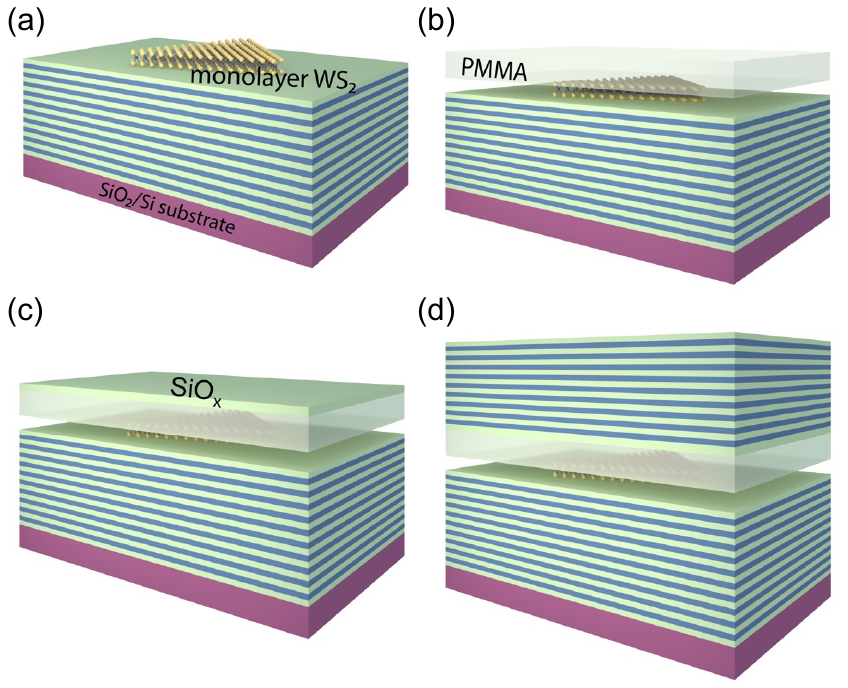}
\caption{\textbf{Design and fabrication procedure for the microcavity.}  (a-d) Schematics of the fabrication steps starting from (a) assembly of the base structure consisting of a  TiO$_2$/SiO$_2$ DBR substrate and a monolayer WS$_2$, followed by (b) the deposition of the PMMA and (c) SiO$_x$ layers, and finished by (d) the deposition of the top SiN$_x$/SiO$_x$ DBR via PECVD.}
\label{fig:Fig1}
\end{figure}
The fabrication procedure for the PMMA/SiO$_x$ spaced planar microcavities tested and discussed in this work is schematically demonstrated in Fig.~\ref{fig:Fig1}a-d. 
The  TiO$_2$/SiO$_2$ DBR substrate, which consists of 17.5 mirror pairs, is fabricated via sputtering \cite{DBR} and finishes with a SiO$_2$ capping layer that has a thickness corresponding to half of the cavity spacer. After transferring the monolayer onto the substrate (see Fig.~\ref{fig:Fig1}a), a PMMA layer is spin-coated on top, which acts both as a protection layer and a cavity spacer (see Fig.~\ref{fig:Fig1}b).
The remaining part of the cavity-spacer (SiO$_x$) and the top DBR (SiN$_x$/SiO$_x$), starting with the SiN$_x$ layer, are deposited via PECVD at 150 $^{\circ}$C to finalise the microcavity (see Fig.~\ref{fig:Fig1}c-d and Supplementary Section S2 for details). The SiO$_x$ spacer allows us to fine-tune the cavity energy and also prevents cracking of the top DBR, which we found to occur when directly depositing it onto PMMA. Most effective confinement of photons with the wavelength $\lambda_C$ is achieved when the total cavity spacer and each DBR layer have a thickness of $\lambda_C/2n$ and $\lambda_C/4n$, respectively, where $n$ is the respective refractive index of each layer \cite{Microcavities}. In principle, the DBR substrate can be fabricated with any deposition technique suitable for making highly reflective DBRs with a smooth, high-quality layer finish.

\begin{figure}
\centering
\includegraphics[width=8.6cm]{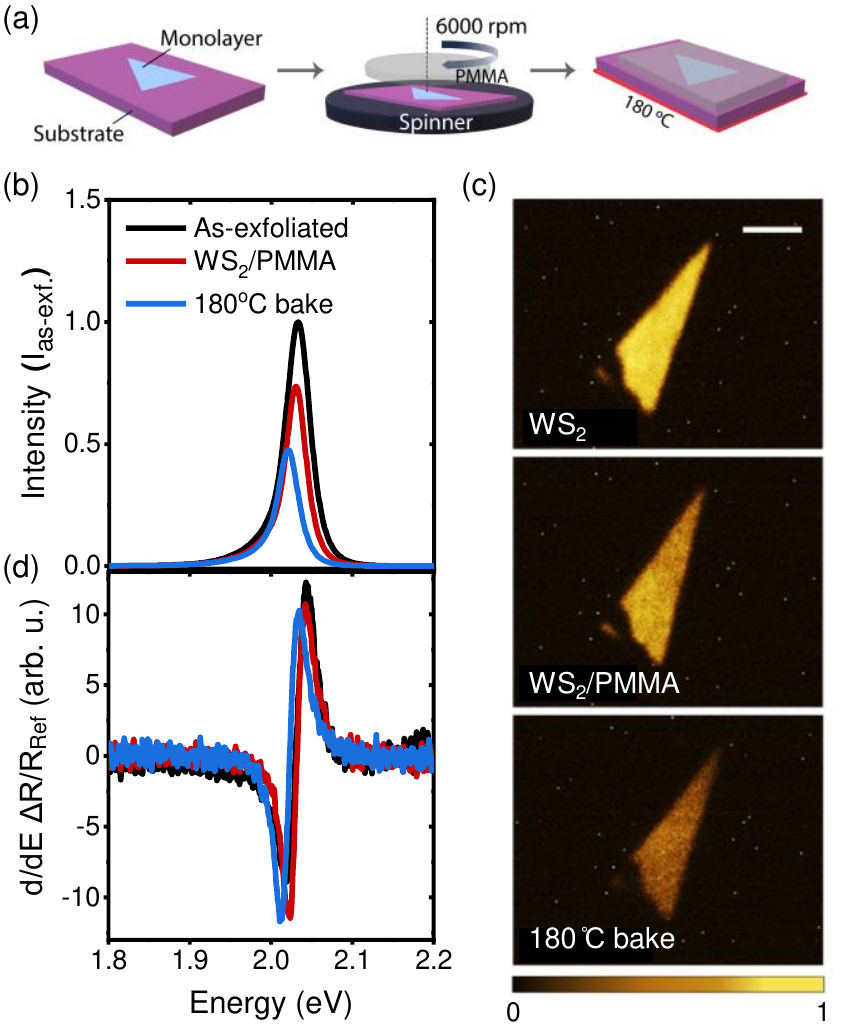}
\caption{{\textbf{Effects of PMMA on the optical properties of WS$_2$.} (a) Schematics of the PMMA encapsulation process. (b-c) Comparison of (b) the PL spectra, (c) the corresponding PL maps, and (d) the derivative of the reflectance contrast spectra for the as-exfoliated monolayer, monolayer capped by PMMA, and the PMMA-capped monolayer baked at 180 $^{\circ}$C. The scale bar size in panel (c) is $10~\mu\mathrm{m}$.}}
\label{fig:Fig2}
\end{figure}

To understand the effect of PMMA deposition on the optical properties of monolayer WS$_2$ (see Fig. \ref{fig:Fig1}b), we mechanically exfoliated a monolayer (bulk crystals sourced from HQ Graphene~\cite{HQgraphene}) on commercial SiO$_2$/Si substrates (sourced from Nova materials~\cite{Novamaterials}). A $80~\mathrm{nm}$ layer PMMA (molecular weight $ \sim $ 950,000 g/mol dissolved in anhydrous anisole with a 2.2\% concentration) was then spin-coated on top of the monolayer at 6000 rpm for 60 s (see Fig.~\ref{fig:Fig2}a). To fully remove the anisole residues and ensure a high-quality surface finish, the PMMA was hard baked at 180 $^{\circ}$C for 90 s~\cite{lundt2016room}.

The tests of the optical response of the WS$_2$ excitons were performed via real-space photoluminescence (PL) and white-light reflectivity measurements under ambient conditions. For the PL studies, we excited the samples with a large Gaussian spot ($d\approx 25~\mu\mathrm{m}$) from a frequency doubled ND:YAG continuous wave (cw) 532-nm laser source with the excitation energy (E$\approx$ 2.33 eV~\cite{zeng2013optical}) above the A-valley exciton energy of monolayer WS$_2$ \cite{Mueller2018}. The reflectivity spectra were measured under illumination of the sample surface with a tungsten halogen white light source. The PL spectra were obtained by averaging the signal from a small sample area, and the derivative of the reflectance contrast spectra $d(\Delta R/R_{\rm ref})/dE$ was derived by also measuring the white light reflection next to the monolayer as reference $R_{\rm ref}$, with $\Delta R=R-R_{\rm ref}$. Since the product of linewidth and amplitude of the reflectance contrast spectrum at the excitonic resonance scales with the exciton oscillator strength, it quantifies the light-matter interaction in the monolayer \cite{Lundt2016,Schneider2018} required for the strong exciton-photon coupling regime. 

The PL spectra in Fig.~\ref{fig:Fig2}b show the PL quenching after the capping of monolayer WS$_2$ with PMMA, and further degradation after baking it. This change is also seen in the corresponding PL maps (see Fig.~\ref{fig:Fig2}c). The pronounced PL quenching effect is possibly due to the chemical instability of this material. WS$_2$ crystals are normally n-type semiconductors due to the presence of sulphur vacancies~\cite{Junqiao2013,Shang2015,carozo2017optical,Sebait2021}, and therefore, they can be strongly affected by the surrounding molecules~\cite{Tongay2013}, such as organic and water molecules that can act as dopants~\cite{zhang2019carbon}. As a consequence, the PMMA capping process can modify the doping level of the monolayer effectively causing a reduction of the exciton quantum yield \cite{Tongay2013}. The additional decrease of the exciton PL after the baking process is likely caused by the elevated temperature accelerating the aging process (oxidation) of the monolayer, which further decreases the exciton quantum yield~\cite{gao2016aging}. However, the derivative of the reflectance contrast spectra in Fig.~\ref{fig:Fig2}d reveals that the exciton oscillator strength is barely affected after the PMMA capping and baking, which indicates that PMMA encapsulation will have a negligible effect on the exciton-photon interactions.

\begin{figure}
\centering
\includegraphics[width=8.6 cm]{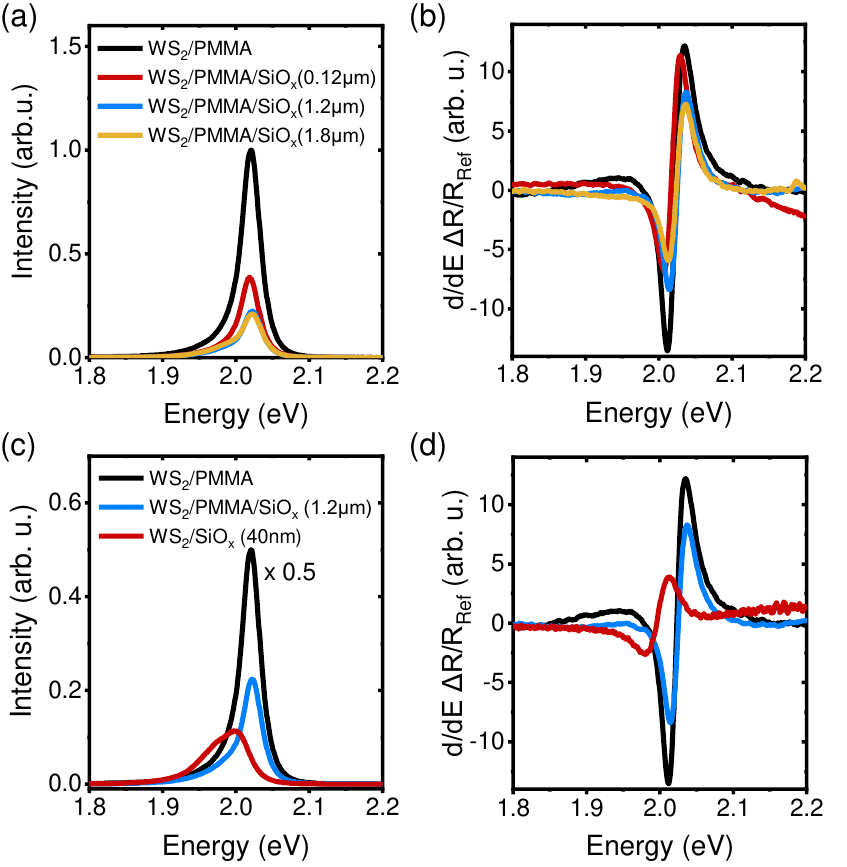}
\caption{\textbf{Effects of PECVD deposition on the optical properties of WS$_2$ with and without PMMA passivation.} (a-b) Comparison of  (a) the PL and  (b) the derivative of the reflectance contrast spectra for monolayer WS$_2$ capped with PMMA, and further overgrown with SiO$_x$ of different thicknesses by PECVD. (c-d) Comparison of (c) the PL spectra and (d) the derivative of the reflectance contrast spectra of monolayer WS$_2$ after deposition of (black) PMMA, (blue) PMMA/SiO$_x$ and (red) SiO$_x$.}
\label{fig:Fig3}
\end{figure}

Since the monolayers are fragile and easily damaged by many fabrication techniques including PECVD~\cite{yun2021influence}, we further tested how well the PMMA layer protects the monolayer against PECVD of SiO$_x$. The influence of the deposition on the exciton PL and absorption in the PMMA-capped monolayer WS$_2$ is demonstrated in Fig.~\ref{fig:Fig3}a-b. Clearly, the exciton PL intensity significantly decreases after the deposition process. The observed quenching is stronger with a larger SiO$_x$ thickness, which scales with the duration of the deposition. Between the layer thicknesses of 1.2 $\mu$m SiO$_x$ and 1.8 $\mu$m SiO$_x$ the PL intensity remains constant and yields $ \sim 25\%$ of the PL intensity of the PMMA-capped WS$_2$, which shows that PECVD induced PL quenching saturates at this level. Similarly, the exciton absorption shown in Fig.~\ref{fig:Fig3}b also decreases after material deposition. This degradation of the PL and oscillator strength could be attributed to the penetration of gaseous plasma (leakage of conductive ions) through the microporous PMMA layer~\cite{charlot2011pvdf}. Once the plasma reaches the monolayer surface, it can degrade the material by creating additional defects, which explains the observed behaviour. Nevertheless, the decrease of excitonic absorption after material deposition is much less compared to that of the excitonic PL. Accordingly, the oscillator strength remains relatively high after deposition of PMMA and SiO$_x$ in contrast with the exciton quantum yield.

To further test the protective properties of the PMMA layer, we prepared an uncapped monolayer sample, on which we directly deposited $40~\mathrm{nm}$ of SiO$_x$ via PECVD. This thickness was chosen to maintain some optical response from the exciton in the WS$_2$/SiO$_x$ structure \cite{yun2021influence}. The PL spectra in Fig.~\ref{fig:Fig3}c reveal that the PL emission of PMMA-capped monolayers with subsequent 1.2 $\mu$m SiO$_x$ deposition is almost twice as strong as that of the uncapped monolayer after deposition of 40 nm SiO$_x$. The reflectance contrast spectra (see Fig.~\ref{fig:Fig3}d) also show much stronger degradation of the exciton oscillator strength for the uncapped monolayer compared to the capped one. Furthermore, the uncapped monolayer shows a pronounced shoulder in the PL spectrum at $E\approx1.97~\mathrm{eV}$ likely stemming from trion emission \cite{Mak2013,Zhu2015,Shang2015}, which indicates a significant defect-induced charge doping effect caused by the deposition process~\cite{Sebait2021,yun2021influence,Sanchez2018}. These results clearly highlight that the PMMA layer protects the monolayer against the dielectric deposition by PECVD. Since the thickness of deposited SiO$_x$ is directly related to the duration of exposure to plasma, the PMMA-capped monolayer was exposed to the plasma for 30 times longer, but still maintained a larger excitonic response, which confirms the protective function of the PMMA layer.

\begin{figure}
\centering
\includegraphics[width=8.6 cm]{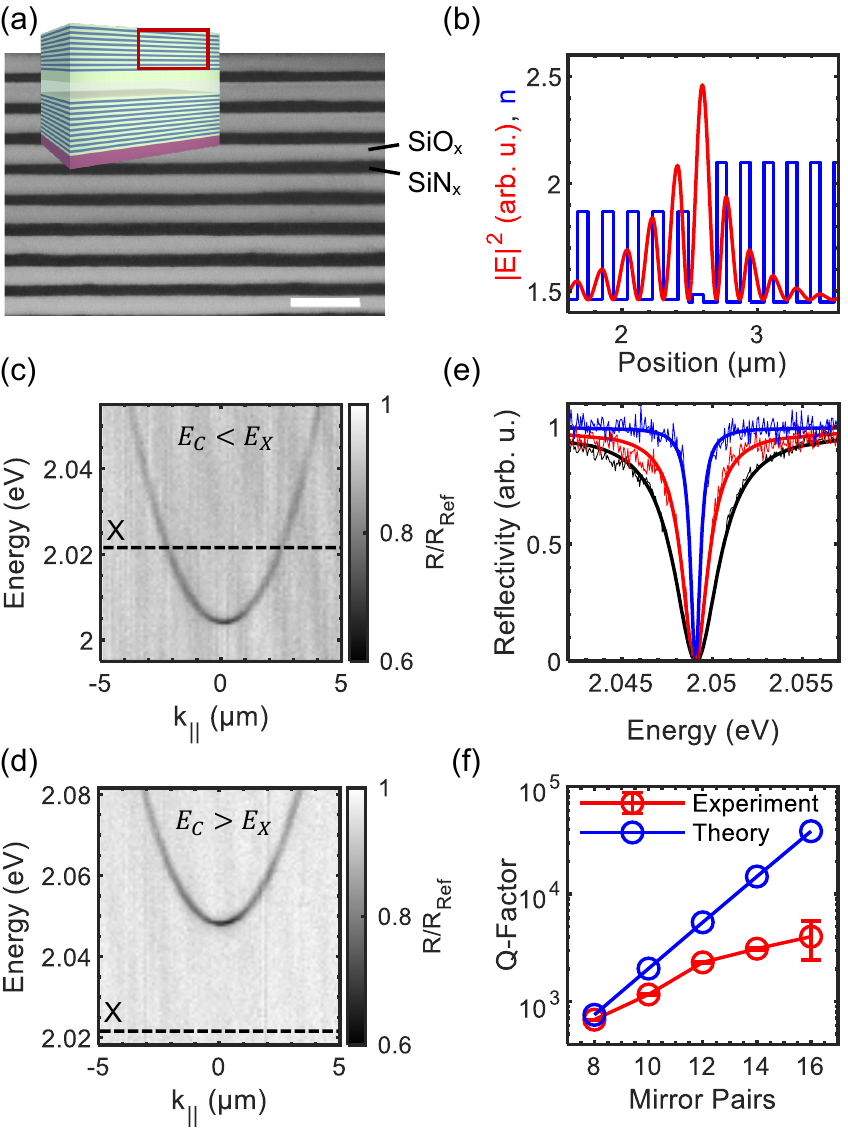}
\caption{\textbf{Design and characterisation of an empty microcavity.}  (a) The SEM image of a SiO$_x$/SiN$_x$ DBR fabricated by PECVD on top of PMMA as marked in (inset) the schematics of the empty microcavity. The scale bar size is $500~\mathrm{nm}$.  (b) The  (blue) refractive indices of the layers around the centre of the microcavity and (red) the simulated E-field distribution of the cavity mode. (c,d)  Angle-resolved reflectivity spectra of microcavities with the resonance energies (c) negatively and  (d) positively detuned from the (X) WS$_2$ exciton energy at room temperature.  (e) Reflectivity spectra at normal incidence with (black) 8.5, (red) 10.5 and (blue) 14.5 SiN$_x$/SiO$_x$ top DBR mirror pairs. (f) Q-factor of the microcavity as a function of top DBR mirror pairs. }
\label{fig:Fig4}
\end{figure}

In addition to testing how PMMA and dielectric material deposition affect the optical response of the WS$_2$ excitons at room temperature, we determined the quality of the photon mode that can be achieved in our cavity without the integration of a monolayer. To ensure that the PMMA/SiO$_x$ cavity provides a suitable substrate for the top DBR, we perform scanning electron microscope (SEM) imaging on the cross-section of a SiN$_x$/SiO$_x$ DBR grown on top of a PMMA layer via PECVD  (see Fig.~\ref{fig:Fig4}a). The SEM image reveals that the SiO$_x$ and SiN$_x$ layers have a consistent thickness throughout the mirror and a smooth lateral profile on the scale of the light wavelength that is crucial for high optical performance.

Figure~\ref{fig:Fig4}b presents the simulation of the optical E-field distribution inside an empty cavity with a cavity-length of $\lambda_C/2n$ fulfilling the Bragg condition for $\lambda_C=615~\mathrm{nm}$, performed with the transfer matrix method \cite{Born2000}. As usual for $\lambda/2$-cavities \cite{Microcavities}, the strongly confined resonant E-field has its maximum near the centre of the cavity, which allows for maximum energy exchange between photons and excitons when the monolayer is placed at this position.  This is ensured by our sample design, where the last layer of the DBR substrate corresponds to the first half of the cavity spacer. By varying the thickness of the SiO$_x$ layer on top of the PMMA layer, both having a similar refractive index, we can fine-tune the cavity length $d$, and therefore, the cavity mode energy E$_{C}$. This, in turn, enables control of the exciton-photon detuning: $\Delta$ = E$_{C}$-E$_X$, where E$_X$ is the exciton energy.

 The angle-resolved reflectivity spectra for empty cavities with $110~\mathrm{nm}$ and $95~\mathrm{nm}$ of PMMA/SiO$_x$ spacer are presented in Fig.~\ref{fig:Fig4}c-d, respectively, with the top DBRs consisting of 13.5 SiN$_x$/SiO$_x$ mirror pairs. The spectra clearly demonstrate the characteristic parabolic dispersion of the cavity photons and show that the photon energy can be tuned well across the exciton energy. The linewidth of the cavity photon ${\rm FWHM}_C$ can be controlled with the number of mirror pairs of the top DBR (see Fig. \ref{fig:Fig4}e), which allows us to increase the quality factor $Q=E_C/{\rm FWHM}_C$ to $Q\approx4\cdot10^3$ reaching the resolution limit of our optical setup (see Fig. \ref{fig:Fig4}f). These results demonstrate that this sample design enables a good control over both the cavity photon energy and linewidth and allows us to reach much higher Q-factors compared to those of the similarly constructed DBR/metal cavities, e.g., \cite{lundt2016room,Gu2019,Wurdack2017} (see Supplementary Table S1). However, to change the exciton-photon detuning, the sample needs to be reproduced with a different thickness of the cavity spacer, and the achievable Q-factors are smaller than the theoretically expected values (see Fig. \ref{fig:Fig4}f). The growth of the Q-factor with the number of mirror pairs in the top DBR decreases above 12 pairs, likely due to scattering and absorption processes in the dielectric layers.
\begin{figure}[htp!]
\centering
\includegraphics[width=8.6cm]{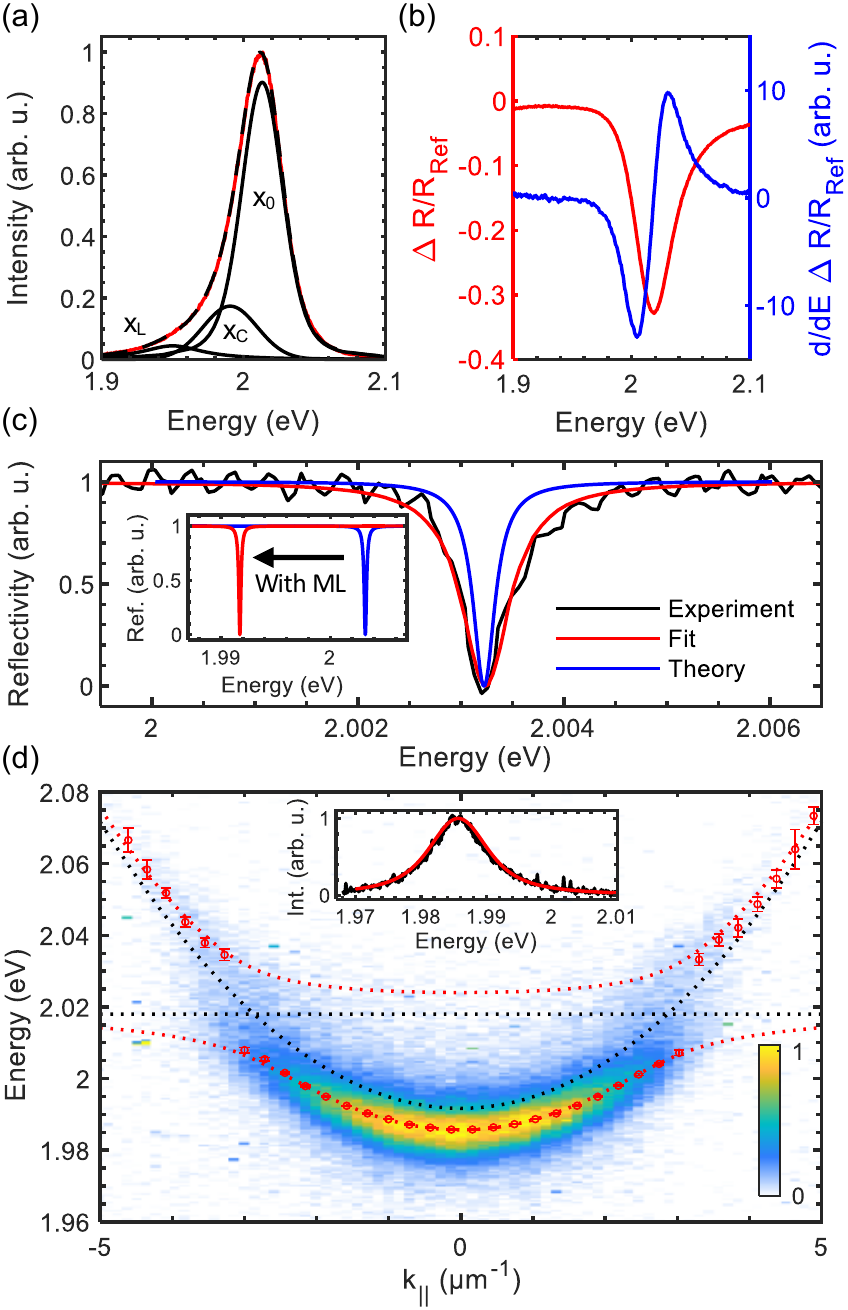}
\caption{\textbf{Characterisation of the microcavity with an integrated monolayer WS$_2$.}  (a) PL spectrum of a monolayer WS$_2$ on top of the DBR substrate fitted with (black) a three peak Voigt function accounting for (X$_\mathrm{L}$) localised, (X$_\mathrm{C}$) charged and (X$_0$) neutral excitons. (b) Reflectance contrast spectrum of the monolayer WS$_2$ and its derivative. (c) Reflectivity spectrum of the cavity around the photon energy at normal incidence next to the position of the monolayer fitted with a (red) Lorentzian function; (blue) theoretical reflectivity spectrum calculated via transfer matrix calculations. (inset) Calculated photon energies at $k_{||}=0$ with and without the monolayer (ML). The cavity energy is shifted by $\sim 11.4~\mathrm{meV}$ due to the optical thickness of the monolayer ($d_{\rm WS_2}\approx0.7~\mathrm{nm}$ \cite{Mueller2018}, $\varepsilon_B^{\rm WS_2}\approx15.831$\cite{Li2014}). (d) Angle-resolved PL spectrum of the full microcavity. The dotted black lines correspond to the experimentally determined exciton and photon energies (see panels a-c), the red dotted lines are the resulting lower and upper polariton branches Eq. \eqref{eq:eigenvalues} with $V=(15.3\pm 0.3)~\mathrm{meV}$ and the red circles are the fitted peak energies (see Supplementary Section S3). (Inset) (black) measured and (blue) calculated PL peak of the lower polariton at $k_{||}=0$.}  
\label{fig:Fig5}
\end{figure}

Finally, we fabricated the whole microcavity with an integrated monolayer WS$_2$, as schematically shown in Fig.~\ref{fig:Fig1}.
Figure~\ref{fig:Fig5}a presents the PL spectrum of a monolayer WS$_2$ placed on the DBR substrate, after deposition of the PMMA spacer (see Fig.~\ref{fig:Fig1}b). 
When fitting the spectrum with a three peak Voigt-function accounting for the localised, charged and neutral exciton emission \cite{Shang2015} (see Fig. \ref{fig:Fig5}a), we can extract the energy and the linewidth of the neutral exciton peak:  $E_X\approx2.013~\mathrm{eV}$ and ${\rm FWHM}_X\approx 35~\mathrm{meV}$, with a contribution of homogeneous broadening ${\rm FWHM}_X^H\approx 12~\mathrm{meV}$ and inhomogeneous broadening ${\rm FWHM}_X^{IH}\approx 28~\mathrm{meV}$. The large spectral weight of the neutral exciton is an evidence of the high quality of the DBR substrate causing negligible doping \cite{Sanchez2018} or strain effects \cite{Khatibi2018,He2013} in the monolayer, which can reduce exciton-photon interactions and prevent strong light-matter coupling.

The reflectance contrast spectrum and its derivative with respect to the energy (see Fig.~\ref{fig:Fig5}b) reveal the exciton energy $E_{X}\approx 2.018~\mathrm{eV}$ and linewidth $FWHM_X\approx35~\mathrm{meV}$, which is in agreement with the Stokes shifted exciton peak of the PL spectrum, and show a significant exciton oscillator strength in the monolayer that ensures a robust exciton-photon interaction strength.

After deposition of the SiO$_x$ spacer and the top DBR consisting of 13.5 SiN$_x$/SiO$_x$ layer pairs, we characterise the cavity mode away from the monolayer area. Figure~\ref{fig:Fig5}c shows the measured and theoretical reflectivity spectra of the microcavity with the cavity mode energy $E_{C}(k_{||}=0)\approx 2.0032~\mathrm{eV}$ and linewidths ${\rm FWHM}_C^{\rm Exp} = (565\pm13)~\mu\mathrm{eV}$ and ${\rm FWHM}_C^{\rm Th} = 257~\mu\mathrm{eV}$, respectively. While the corresponding measured Q-factor $Q^{\rm Exp}\approx3500$ is smaller than the theoretical Q-factor $Q^{\rm Th}\approx 7800$, it is similar to the Q-factor of the empty microcavity (see Fig. \ref{fig:Fig4}f) and in general, comparable to the largest values reached for microcavities with embedded TMDCs \cite{Sidler2017, Rupprecht2021, Knopf2019, Shan2021, Wurdack2021} (see Supplementary Table S1).

Figure~\ref{fig:Fig5}d presents the angle-resolved PL measurement of the whole structure, which was obtained via excitation of the monolayer with a tightly focused laser spot ($\lambda=532~\mathrm{nm}$) and collecting the PL in momentum space with a NA=0.65 objective, projected into a 2D spectrometer. The measured dispersion, which decreases in intensity at larger energies due to thermalisation \cite{lundt2016room,Wurdack2021}, can be approximated with the model of two coupled oscillators \cite{Savona1995}:
\begin{equation}
\label{eq:eigenvalues}
    \bar{{E}}_{U/L} = 0.5\left(\bar{E}_{X}+\bar{E}_C \pm \sqrt{(\Delta-i\delta)^2 + 4V^2}\right),
\end{equation}
where $\bar{E}=E-i\gamma$ are the complex energies of the lossy modes, $V$ is the strength of coupling between the cavity photons and the excitons, and $\delta = \gamma_C-\gamma_X$ (see Fig. \ref{fig:Fig5}d). Note, that $\gamma$ corresponds to ${\rm HWHM}$ of the measured modes \cite{Savona1995}.

Using the energies and linewidths of the exciton and cavity resonances (see Fig. \ref{fig:Fig5}a-c), we find an excellent agreement between Eq. \eqref{eq:eigenvalues} and the measured dispersion with $V=(15.3\pm 0.3)~\mathrm{meV}$ (see Fig. \ref{fig:Fig5}d). The corresponding Hopfield coefficient of the lower polariton branch 
\begin{equation} 
\label{eq:Hopfield}
|X|^2=0.5\left[1+\frac{\Delta}{\sqrt{\mathfrak{Re}\left(\sqrt{4V^2- \delta^2}\right)}
+\Delta^2}\right]
\end{equation}
allows us to approximate the polariton PL spectrum with a Voigt profile, where ${\rm FWHM}_L^{IH/H}=|X|^2 {\rm FWHM}_X^{IH/H} + \left(1-|X|^2\right){\rm FWHM}_C^{IH/H}$, which reproduces the measured PL well (see Fig. \ref{fig:Fig5}d, inset). Since $V>|\delta|/2\approx 8.6~\mathrm{meV}$ and $V>0.5\sqrt{\gamma_X^2+\gamma_C^2}\approx 8.8~\mathrm{meV}$, the sample operates in the strong exciton-photon coupling regime, exhibiting  distinct exciton-polariton modes \cite{Savona1995}. Hence, the sample fabrication technology presented here is suitable for making high-Q planar microcavities with an integrated atomically-thin WS$_2$ crystal operating in the strong light-matter coupling regime. 

In summary, we have developed a scalable approach for making high-quality planar microcavities with integrated atomically-thin semiconductors suitable for hosting exciton-polaritons at room temperature. This is achieved by depositing the two DBRs, a monolayer WS$_2$ and a PMMA/SiO$_x$ spacer layer-by-layer. We demonstrate that the PMMA layer has only a minor effect on the WS$_2$ exciton oscillator strength and effectively protects the monolayer against further deposition of the top DBR, which allows for strong light-matter interactions in the structure. Since the PMMA/SiO$_x$ spacer and the top DBR cover the whole microchip homogeneously, the functional area of the loaded microcavity is mainly limited by the size of the monolayer, which can be scaled up to cm-scale with recently developed synthesis \cite{Lee2017,Choi2022} and exfoliation \cite{Desai2016,Fang2020} techniques. Further, the PMMA can be easily patterned by lithography methods, paving the way towards creating polariton lattices, waveguides and other 2D potential landscapes on a microchip, as previously demonstrated for microcavities with embedded perovskites \cite{Su2020,Su2018}.
\section*{Acknowledgements}
\begin{acknowledgments}
This work was supported by the Australian Research Council (ARC) grants CE170100039 and DE220100712. We acknowledge the Australian National Fabrication Facility (ANFF) OptoFab at its node in the Australian Capital Territory (ACT) for fabricating the DBR substrate and technical support for sample fabrication from the ANFF ACT node.
\end{acknowledgments}

\bibliography{bibliography}

\begin{thebibliography}{64}
\bibitem [{DBR()}]{DBR}%
  \BibitemOpen
  \href@noop {} {\bibinfo  {journal} {The fabrication of the high-reflectivity
  substrate was peformed at the ACT OptoFab node of the Australian National
  Fabrication Facility - a company established under the National Collaborative
  Research Infrastructure Strategy to provide nano and microfabrication
  facilities for Australia’s researchers}\ }\BibitemShut {NoStop}%
\end{thebibliography}

\end{document}



\title{Supplementary Material: \\ Fabrication of high-quality PMMA/SiO$_x$ spaced planar microcavities for strong coupling of light with monolayer WS$_2$ excitons} 

\author{Tinghe~Yun}%
\affiliation{ARC Centre of Excellence in Future Low-Energy Electronics Technologies and Department of Quantum Science and Technology, Research School of Physics, The Australian National University, Canberra, ACT 2601, Australia}
\affiliation{Songshan Lake Materials Laboratory, Dongguan 523808, Guangdong, China} 
\affiliation{Institute of Physics, Chinese Academy of Science, Beijing, 100190, China}

\author{Eliezer~Estrecho}%
\affiliation{ARC Centre of Excellence in Future Low-Energy Electronics Technologies and Department of Quantum Science and Technology, Research School of Physics, The Australian National University, Canberra, ACT 2601, Australia}

\author{Andrew~G.~Truscott}%
\affiliation{Department of Quantum Science and Technology, Research School of Physics, The Australian National University, Canberra, ACT 2601, Australia}

\author{Elena~A.~Ostrovskaya}
\email{elena.ostrovskaya@anu.edu.au}
\affiliation{ARC Centre of Excellence in Future Low-Energy Electronics Technologies and Department of Quantum Science and Technology, Research School of Physics, The Australian National University, Canberra, ACT 2601, Australia}

\author{Matthias~J.~Wurdack}
\email{matthias.wurdack@anu.edu.au}
\affiliation{ARC Centre of Excellence in Future Low-Energy Electronics Technologies and Department of Quantum Science and Technology, Research School of Physics, The Australian National University, Canberra, ACT 2601, Australia}



\date{\today}

\pacs{}

\maketitle 
 \onecolumngrid

\section*{S1: Summary of fabrication procedures and Q-factors of TMDC-based microcavities}
 \begin{table*}[htp!]

\begin{center}
    \begin{tabular}{| l | l | l | l | l | }
    \hline 
    {\bf Cavity Architecture} & {\bf Spacer Material} & {\bf Fabrication Technique} & {\bf Q-factor} & {\bf Ref} \\ \hhline{|=|=|=|=|=|} 
        \makecell[l]{
        Open microcavity with \\ 
            a concave cavity spacer} & 
        Air gap & 
        DBR deposition on separate substrates. &
        $\sim 4200$ & 
        \cite{Sidler2017}  
        \\ \hline 
        Open microcavity & 
        Air gap &  
        DBR deposition on separate substrates. &
        $\sim 160$
        & \cite{Krol2019} 
        \\ \hline
        All-dielectric microcavity & 
        SiO$_2$ &  
        Bottom-up deposition by PECVD. &
        $\sim 250$ & 
        \cite{Liu2014,Chen2017} 
        \\ \hline
        DBR/metal cavity & 
        SiO$_2$/h-BN/PMMA & 
        \makecell[l]{Bottom-up deposition of DBR by PECVD, hexagonal \\ 
            Boron Nitride (hBN) by mechanical transfer, PMMA spacer \\ 
            by  spin-coating and  top metal mirror by electron beam \\ 
            evaporation (EBE).} &
        $\sim 78$ & 
        \cite{Gu2019} 
        \\ \hline
        DBR/metal cavity  & 
        SiO$_2$/PMMA & 
        \makecell[l]{Bottom-up deposition of DBR by sputtering, PMMA \\ 
            spacer by spin-coating and top metal mirror by EBE.} &
        $\sim 110$ & 
        \cite{lundt2016room} 
        \\ \hline
        DBR/metal cavity  & 
        AlAs/GaInP/PMMA & \makecell[l]{Bottom-up deposition of DBR and cavity spacer by \\ 
            molecular beam epitaxy (MBE), PMMA spacer by spin- \\
            coating and top metal layer by EBE.} &
        $\sim 650$ & 
        \cite{Wurdack2017} 
        \\ \hline
        All-dielectric microcavity  & 
        SiO$_2$/HSQ/Al$_2$O$_3$ & 
        \makecell[l]{Bottom-up deposition of DBR substrate by PECVD, \\ 
            hydrogen silsesquioxane (HSQ) spacer by spin-coating, \\
            Al$_2$O$_3$ spacer by  atomic-layer deposition (ALD) and \\
            top DBR by PECVD.} & 
            $\sim 600$ & \cite{Liu2017,Liu2020} \\ \hline
        All-dielectric microcavity 
        & hBN/SiO$_2$ & 
        \makecell[l]{Bottom-up deposition of DBR substrate by ion-assisted \\ 
        physical vapor deposition (IAD), hBN via mechanical \\ 
        transfer, and SiO$_2$ spacer and top  DBR via IAD.} & 
        $\sim 4500$ & 
        \cite{Knopf2019,Shan2021} 
        \\ \hline
        ``Flip-Chip" microcavity & 
        PMMA/SiO$_2$ & 
        \makecell[l]{Deposition of DBR substrate by sputtering, PMMA spacer \\ 
            by spin-coating and mechanical transfer of the top DBR.} & 
        $\sim 4500$ & 
        \cite{Lundt2019,Rupprecht2021} 
        \\ \hline
        \makecell[l]{``Flip-Chip" all-dielectric \\ 
            microcavity} & 
        SiO$_2$ & 
        \makecell[l]{Deposition of DBR substrate by PECVD and sputtering, \\ 
            and mechanical transfer of the top DBR.} &
        $\sim 3000$ & 
        \cite{Rupprecht2021,Wurdack2021} 
        \\ \hline
        This work & 
        PMMA/SiO$_x$ & 
        \makecell[l]{Bottom-up deposition of DBR substrate by sputtering, \\ PMMA spacer by spin-coating, and SiO$_x$ spacer and top \\DBR by PECVD.} &
        $\sim 3500$ & 
        \\ \hline
\end{tabular}
\caption{Summary of fabrication procedures and Q-factors of planar microcavities with integrated monolayer TMDCs operating in the strong exciton-photon coupling regime.}
\end{center}

 \end{table*}
 \newpage
 
 \twocolumngrid
\section*{S2: Deposition of the silicon oxide spacer and the top DBR}
The SiO$_x$ spacer on top of the PMMA, and the top SiN$_x$/SiO$_x$ DBR were deposited  via plasma enhanced chemical vapour deposition (PECVD) using the Oxford PlasmaLab System 100 operating at radio frequency (RF). The pressure of the PECVD chamber was set to $p=650~\mathrm{mTor}$, the temperature to $T=150~\mathrm{^\circ C}$ and the power of the RF generator to $P=30~\mathrm{W}$. For the SiO$_x$ deposition, we introduced $161~\mathrm{sccm}$ of N$_2$ gas, $710~\mathrm{sccm}$ of N$_2$O gas and $9~\mathrm{sccm}$ of SiH$_4$ gas, and for the SiN$_x$ deposition, $980~\mathrm{sccm}$ of N$_2$ gas, $14~\mathrm{sccm}$ of NH$_3$ gas and $18~\mathrm{sccm}$ of SiH$_4$ gas into the deposition chamber. The growth rates and the refractive indices of the deposited SiN$_x$ and SiO$_x$ layers were initially determined via ellipsometry, ensuring that the layers meet the Bragg condition for light with the wavelength of the WS$_2$ excitons, $\lambda_C\approx\lambda_X$.
 
\section*{S3: Cross section of the angle-resolved PL spectrum}
\begin{figure}
\centering
\includegraphics[width=8.6cm]{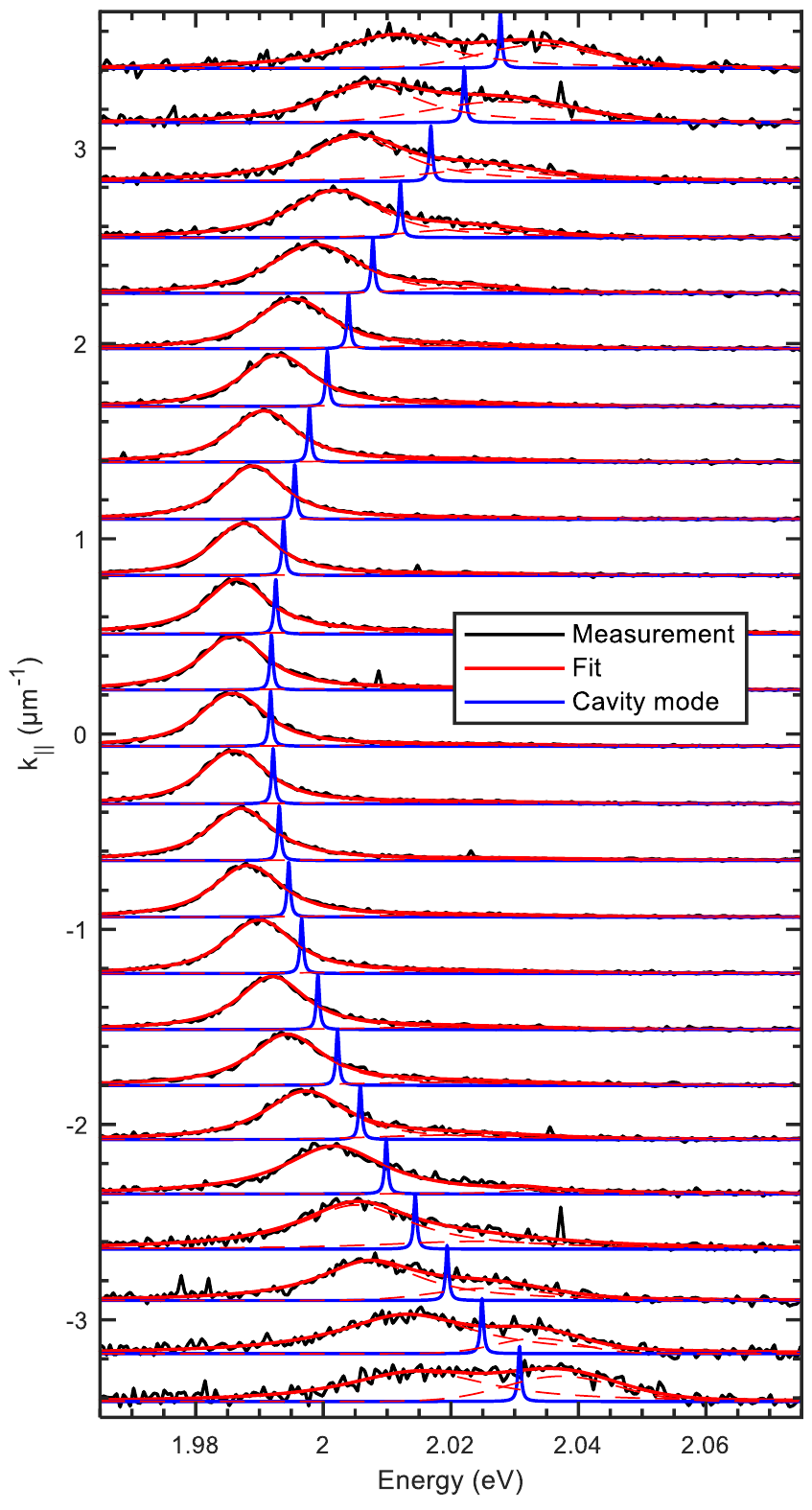}
\caption{Normalised PL spectra extracted from the angle-resolved PL measurement of the loaded microcavity (see manuscript Fig. 5d), fitted with two-peak Voigt functions, where the baseline of each spectrum marks its corresponding in-plane momentum. The cavity mode at the position of the WS$_2$ monolayer is plotted as a reference.}
\label{fig:Fig1-SI}
\end{figure}
To extract the peak energies of the angle-resolved PL spectrum of the loaded microcavity (shown in the manuscript Fig. 5d), we fitted to each spectrum shown in Fig. \ref{fig:Fig1-SI} a two-peak Voigt function, which are plotted together with the cavity mode on the position of the monolayer as reference. The fitted peak positions are marked in Fig. 5d and follow well the expected lower and upper polariton branches in this system.\\
\\
\bibliography{bibliography-SI}